# Very Massive Close Binaries and the Puzzling Temporal Evolution of $^{14}$N in the Solar Neighbourhood

*D. Vanbeveren and E. De Donder*

Astrophysical Institute, Vrije Universiteit Brussel, Pleinlaan 2, 1050 Brussels, Belgium

dvbevere@vub.ac.be

**Abstract**

Low metallicity very massive stars with an initial mass between 140 Mo and 260 Mo can be subdivided into two groups: those between 140 Mo and 200 Mo which produce a relatively small amount of Fe, and those with a mass between 200 Mo and 260 Mo where the Fe-yield ejected during the supernova explosion is enormous. We first demonstrate that the inclusion of the second group into a chemical evolutionary model for the Solar Neighbourhood predicts an early temporal evolution of Fe which is at variance with observations whereas it can not be excluded that the first group could have been present. We then show that a low metallicity binary with very massive components (with a mass corresponding to the first group) can be an efficient site of primary $^{14}$N production through the explosion of a binary component that has been polluted by the pair instability supernova ejecta of its companion.

When we implement these massive binary $^{14}$N yields in a chemical evolution model, we conclude that very massive close binaries may be important sites of $^{14}$N enrichment during the early evolution of the Galaxy.

**Keyword**

Galaxies: evolution -- Binaries: evolution

## 1. Introduction

To study the chemical evolution of a galaxy, one generally adopts the cosmological picture in which it is assumed that all deuterium, a major part of helium and some lithium have been produced during the Big Bang whereas all the heavier elements (commonly referred to as metals) are made in stars. Following the overall picture proposed originally by Tinsley (1980), a first generation of metal free stars forms from the collapsing protogalaxy. They process the matter nuclearly and return it partially or completely to the interstellar medium (ISM) via stellar winds (SWs) and/or supernova (SN) explosions. Out of this enriched gas a new generation of stars is born which further enrich the ISM. The cyclic process of stellar birth and death continues until no more gas is left to create new stars. The first metal-free stars (Population III stars) should have distributed according to an initial mass function (IMF) more skewed towards massive stars than today (see, e.g. Abel et al., 2002; Bromm and Larson, 2004). The ensemble of a galaxy formation model, the dynamics of stars and gas clouds in a galaxy, a star formation model and the details of stellar evolution from birth till death as function of chemical composition makes the chemical evolutionary model (CEM) of a galaxy. During the last four decades many groups all over the world have constructed CEMs with varying degree of sophistication. When we restrict ourselves to the last couple of years and to the early evolution of our own Galaxy, the following list may be representative (Sommer-Larsen et al., 2003; Argast et al., 2004; Prantzos, 2005; Chiappini et al., 2005, 2006; Ballero et al., 2006; Akerman et al, 2004 ; De Donder and Vanbeveren, 2004 further DV) and references therein. CEMs try to predict the temporal evolution of the α-elements (i.e. C, N, O, Mg, Si, S, Ca), the Fe-peak elements (i.e. Cr, Mn, Fe, Co, Ni, Cu, Zn), s-process elements (i.e. Ba) and r-process elements (i.e. Eu). CEM-predictions depend on parameters and uncertainties of astrophysical processes, not in the least of processes related to stellar evolution and stellar death (the SN explosion). Fortunately, past and present stellar abundance observational projects (one of the most recent is ESO's Large Program "First Stars") allow us to constrain many parameters and uncertainties.

The nuclear reactions which determine the $^{14}$N rates are well known and therefore one would expect that a CEM should have no problem to explain its galactic temporal evolution. However most of the CEM-scenarios published in the past had difficulties to approach the observed $^{14}$N abundance pattern of the Solar Neighbourhood. The latter is in particular true for the early evolutionary phases when the overall metallicity Z was small (say less than $10^{-5}$). To illustrate this foregoing statement we refer to the CEM-simulations of DV where chemical yields are considered resulting from evolutionary calculations of non-rotating massive stars. Meynet and Maeder (2002) investigated the effects of rotation on stellar evolution in general, massive stars in particular. They computed $^{14}$N yields for a population of massive stars all rotating with a zero age main sequence velocity of 300 km/s under the assumption that all massive stars eject most of their mass outside the CO-core before or during the final collapse of the core. Meynet and Maeder argue that starting with an initial 300 km/s corresponds with the observed average rotational velocity of 200 km/s for all O-type stars in the Solar neighbourhood. Prantzos (2005) and Chiappini et al. (2005) implemented these yields into their CEM-codes and concluded that rotation with velocities that are presently observed can not explain the $^{14}$N discrepancy.

If 300 km/s cannot do the job, may be higher values can. First notice that stellar structure calculations illustrate the expected result that the smaller the metallicity the more compact is a star. When stars begin their evolution on the ZAMS with the same angular momentum content, regardless of their metallicity, then a ZAMS value of 300 km/s at Solar metallicity corresponds to 600-800 km/s at $Z = 10^{-8}$ (Meynet et al., 2005) and massive star evolutionary computations with such high rotational velocities then indeed reveal a significant $^{14}$N increase in the stellar interior. The chemical yields published by the Geneva group have been implemented in a CEM (Chiappini et al., 2006) and the results illustrate that indeed very rapidly rotating massive stars that lose a lot of mass prior to or during core collapse can help to solve the $^{14}$N discrepancy. A few words of caution are appropriate here, e.g. let us critically evaluate the statement that massive stars in the Solar neighbourhood have an average rotational velocity of 200 km/s. The rotational velocity distribution of O and early B-type stars has been investigated by Penny (1996) and by Vanbeveren et al. (1998). It was concluded that the distribution is highly asymmetrical with an extended tail towards large values. This means that it is not correct in order to use average rotational velocity values as basis of strong conclusions. The distribution illustrates that there is a majority of relatively slow rotators (with an average velocity of 100 km/s) and a minority of stars that are rotating very fast (statistically on could call them outliers). The tail obviously demonstrates that there are stars which are rapid rotators. Binary mass gainers (runaways like ζ Pup), binary mergers and stellar collision products in young dense stellar environments are expected to be rapid rotators and thus are expected to belong to the tail. The question however is whether or not one can approximate their evolution with rotating single star models.

Due to the dynamo effect, rotation generates magnetic fields (Spruit, 2002) which means that the evolutionary effect of rotation cannot be studied separately from the effects of magnetic fields. This was done recently by Maeder and Meynet (2004) and (as could be expected) several of the stellar properties (size of the core, main sequence lifetime, tracks in the HR diagram, surface abundances) are closer to those of models without rotation than with rotation only. The authors argued that since single star evolution with rotation and magnetic fields does not explain the surface CNO chemistry of the observed massive supergiant population, magnetic fields must be unimportant. However, this argumentation is based on the assumption that most of the massive stars evolve as single stars do. There are however a number of other processes which may be responsible for altered CNO abundances in OB-type stars (the Roche lobe overflow process in interacting binaries where the surface layers of the mass loser but also of the mass gainer may become nitrogen enriched, the merger of two binary components due to a common envelope phase and last not least, the collision and merger process due to N-body dynamics in young dense stellar systems) and before an argumentation as the one of Maeder and Meynet has any meaning, one has to consider all these processes.

All in all, it can be concluded that the $^{14}$N discrepancy during the early evolution of the Milky Way can be reduced provided that the majority of the massive stars at low Z rotate very fast, the star loses a lot of mass prior or during the core collapse and the magnetic fields in rapidly rotating stars are very small.

Is there an alternative way to explain the $^{14}$N discrepancy. In most of the CEMs it is assumed that the presently observed maximum stellar mass of stars in the Solar Neighbourhood (=100-120 Mo) was the same in the past, i.e. independent from Z. The results of the Wilkinson Microwave Anisotropy Probe (WMAP) provide observational evidence that the low metallicity stellar initial mass function was top heavy with the possible presence of stars which were significantly more massive than 100-120 Mo (the term *very massive star* is used). The optical depth along the line of sight to the last scattering surface of the Cosmic Microwave Background is interpreted in terms of the existence of a population of very massive stars in the early Universe (Kogut et al., 2003; Sokasian et al., 2003). In the present paper we will investigate whether the problem of the production of nitrogen in the early Galaxy when the metallicity was small, is related to the possibility that very massive stars were present during this phase.

**2. The formation of very massive low metallicity stars**

How massive stars form is still poorly understood (Stahler et al. 2000; Larson 2003). There are two competing theories (for a detailed discussion see Bally and Zinnecker, 2005). The standard model assumes that massive stars form similar to low-mass stars but with higher accretion rates. A major difference is that massive stars (with a mass >10 Mo) have a pre-main-sequence contraction phase which is shorter than the accretion timescale, which means that massive stars reach the zero age main sequence while they are still accreting. Therefore, the most important difficulty in forming stars with a mass larger than 10 Mo is the radiation pressure which may be sufficient to reverse the infall of gas when it is metal rich and contains typical dust properties (Wolfire and Casinelli, 1987; Beech and Mitalas, 1994). A detailed modeling of star formation through accretion was recently made by Yorke and Sonnhalter (2002) who demonstrated that direct accretion of non-zero metallicity gas can produce stars up to 20-30 Mo. However, observations tell us that stars with a mass significantly larger than 20-30 Mo exist also in high metallicity environments like the Solar Neighbourhood and an alternative massive star formation theory seems to be required. It is a fact that massive stars tend to form in rich clusters rather than in isolation (Clarke et al., 2000; Lada and Lada, 2003, de Wit et al., 2004) and this implies that star formation may be affected by N-body interactions of stars and protostars. The present supercomputers allow to study N-body interactions of massive stars in detail (Portegies Zwart et al. 1999, 2002 for recent simulations) and one of the conclusions related to the process of massive star formation is that physical collisions and merging of stars and protostars may occur very frequently. Bonnell et al. (1998) proposed that massive stars form by a combination of gas accretion and merging of stars and protostars in forming star clusters (see also Bally, 2002; Bonnell, 2002; Zinnecker and Bate, 2002; Boss, 1996; Klessen and Burkert, 2000; Bate et al., 2003; Bonnell and Bate, 2002; Bonnell et al., 2003). These studies allow to conclude that (see also Bonnell, 2005 for a review)

> *massive (and very massive) stars originate in the cores of clusters where due to stellar dynamics they form close binaries with a fairly high mass ratio.*

### 3. The evolution of massive and very massive low metallicity stars

Marigo et al. (2001, 2003) presented evolutionary tracks up to central carbon ignition of zero-metallicity single stars with an initial mass between 0.7 Mo and 1000 Mo. Fryer et al. (2001) and Heger and Woosley (2002) calculated the evolution till the collapse of the Fe-core and the possible supernova explosion of massive low metallicity single star He-cores. As far as the very massive stars is concerned the emerging picture is the following: at the end of their evolution stars with an initial mass larger than 260 Mo and those with a mass between 40 Mo and 140 Mo directly collapse into a black hole without ejecting matter. Stars with an initial mass between 140 Mo and 260 Mo end their life in a pair instability supernova where the star is completely disrupted. From the chemical yields of stars in this mass range tabulated by Heger and Woosley (2002), we restrain the following: the ejecta contain typically $1-4.10^{-5}$ Mo of $^{14}$N which is of primary origin, 4-5 Mo of $^{12}$C, 40-50 Mo of $^{16}$O, 3-4 Mo of $^{24}$Mg, 10-25 Mo of $^{28}$Si, 3-10 Mo of $^{32}$S. Of particular importance are the predicted $^{56}$Fe ejecta: within the mass range 140-200 Mo these ejecta are relatively small, however for larger masses, they become large to very large reaching about 40 Mo for the 260 Mo star.

The CEM-simulations presented here were made assuming that very massive stars formed only when $Z < Z_{max}$ and that their evolution resembles the one summarized above. We investigated the consequences when different values of this maximum are used, namely $Z_{max} = 10^{-6}$ and $Z_{max} = 10^{-5}$.

### 4. The effect of very massive stars on the early galactic chemical evolution

Our CEM which includes a detailed treatment of intermediate mass and massive close binaries has been described in detail in an extended review (De Donder and Vanbeveren, 2004) and references therein. In the present paper, we use the following CEM assumptions.

- To describe the formation and evolution of the Galaxy we use the model discussed by Chiappini et al. (1997) which is supported by recent results of hydrodynamical simulations (Sommer-Larsen et al., 2003).

Stellar evolution:

- The chemical yields of single stars with an initial mass = [0.1 ; 120] Mo are calculated from detailed stellar evolutionary computations where, in the case of massive stars, the most recent stellar wind mass loss rate prescriptions are implemented in the evolutionary code. Of course, the stellar evolutionary dataset depends on the initial metallicity Z and as far as massive stars are concerned not in the least on the effects of Z (read Fe) on the stellar wind mass loss rates. Our dataset corresponds to the case where the Luminous Blue Variable mass loss rate is independent from Z whereas the red supergiant and Wolf-Rayet mass loss rates depend on Z as predicted by the radiatively driven wind theory (Kudritzki et al., 1989).

- Similarly as for the single stars, the chemical yields of binary stars with an initial primary mass = ]1 ; 120] Mo are based on an extended data set of detailed stellar evolutionary calculations of binaries where the evolution of both components is followed simultaneously. Obviously for the massive binaries the stellar wind effects are included similarly as in massive single stars.

- The initial binary mass ratio distribution is a Hogeveen type (Hogeveen, 1992).

- The initial binary period distribution is flat in the Log.

- A Salpeter type IMF for single stars and for binary primaries. We assume that the IMF-slope does not vary in time, which is more than sufficient in order to illustrate the basic conclusion of the present paper.

- To simulate the effects of SN Ia, we calculate the SN Ia rate from first population synthesis principles using the single-degenerate (SD) model of Hashisu et al., 1997, 1999) and the double degenerate (DD) model of Iben and Tutukov (1984) and Webbink (1984).

- We adopt an overall initial binary formation frequency of 50% in the mass range givenabove; this frequency is assumed not to vary in time.

- The stable Roche lobe overflow phase in binaries is assumed to be conservative (all mass lost by the loser is accreted by the gainer). Notice that in most of the binaries with an initial primary mass larger than 40 Mo the Roche lobe overflow phase is avoided due to the very large stellar wind mass loss during the Luminous Blue Variable phase of the primary (the LBV scenario as it was originally introduced by Vanbeveren, 1991).

- We assumed maximum ejection efficiency in binaries that evolve through a common envelope and/or a spiral-in phase.

- The effect of the SN explosion on the binary and binary orbital parameters is included by adopting a distribution of SN-asymmetries (expressed as the kick velocity that the compact SN remnant gets depending on the SN-asymmetry) and integrating over this distribution; the distribution of SN-asymmetries is based on the observed space velocity distribution of single pulsars which let us suspect that the distribution is $\chi^2$-like (or Maxwellian but with a tail extending to very large values) with an average kick-velocity = 450 km/s.

- The maximum neutron star mass = 2 Mo (De Donder and Vanbeveren, 2004) although the exact value does not critically affect the basic conclusion of the present paper.

- Black hole formation happens when the initial mass of a single star >25 Mo and >40 Mo for an interacting binary component. Notice that the minimum-mass-difference is due to the fact that the evolution of the helium core in a single star differs from the one of a binary component who lost its hydrogen rich layers due to Roche lobe overflow during hydrogen shell burning on the thermal timescale (for more details see De Donder and Vanbeveren, 2004). From first physical principles it is unclear whether or not BH formation in massive stars is accompanied by significant matter ejection. De Donder and Vanbeveren (2003) investigated the consequences on CEM simulations of the early Galaxy of the latter uncertainty. It was concluded that a CEM where it is assumed that all massive stars with an initial mass larger than 40 Mo form BH's without matter ejection predicts a temporal $^{16}$O evolution which is at variance with observations. These observations are much better reproduced when during the core helium burning a star loses mass by stellar wind (a Wolf-Rayet type wind) which depends on the stellar Fe-abundance as predicted by the radiatively driven wind theory whereas prior to or during BH formation on average 4 Mo of $^{16}$O should be ejected. This is also the model that we use in the present paper. Notice that in the single star mass range 25 Mo and 40 Mo we use the stellar model computations of Woosley and Weaver (1995).

It was straightforward in order to adapt our CEM and to investigate the effects of very massive stars on the chemical evolution of the Galaxy as function of $Z_{max}$. Our simulations allow to put forward the following overall conclusion:

*for $Z_{max} < 10^{-6}$, i.e. when very massive stars existed only when the galactic metal content Z was smaller than $10^{-6}$, their effect on the overall evolution of the galactic α-elements is small and can be neglected.*

The foregoing result is a confirmation of the results of Ballero et al. (2006). In Figures 1-6 we compare the observed temporal evolution of $^{12}$C, $^{16}$O, $^{24}$Mg, $^{28}$Si, $^{32}$S and Ca with the theoretical ones predicted by our CEM where we included the yields of the very massive stars up to 260 Mo from Heger and Woosley (2002) and $Z_{max} < 10^{-5}$. As can be noticed, the correspondence with the observations is very poor (dashed lines). The main reason is the initial enrichment of Fe which is caused by the very massive stars with an initial mass larger than 200 Mo. We then repeated our CEM simulations but when $Z > 10^{-6}$ we ignored the very massive stars with an initial mass larger than 200 Mo. In this way we simulate the effect either of a truncated IMF or of a very large stellar wind mass loss rate for very massive stars with an initial mass larger than 200 Mo. Figures 1-6 also show the results of the latter simulation (solid lines). As can be noticed the early Fe-enrichment is largely suppressed which means that it cannot be excluded that stars with an initial mass as large as 200 Mo may have been present for quite some time in the early Galaxy.

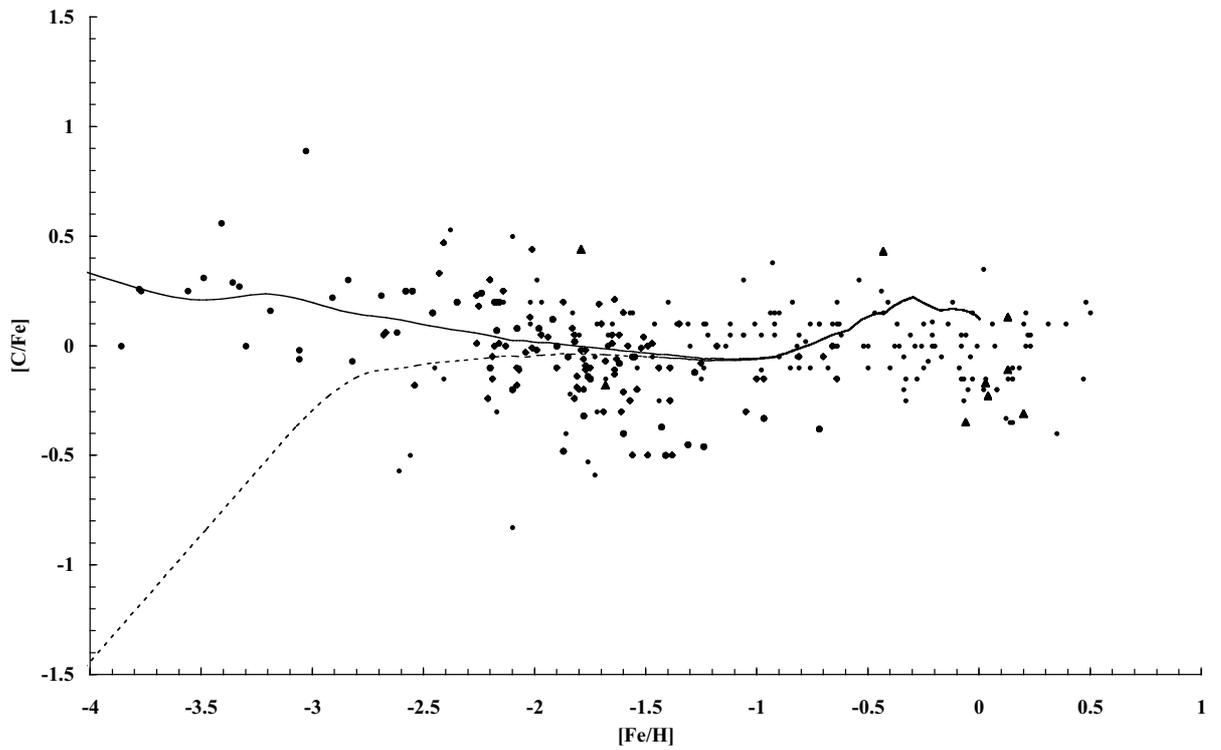

Figure 1: The observed (dots) versus the theoretically predicted temporal evolution of $^{12}$C. The thin line (resp. dashed line) corresponds to the case where very massive single stars with a mass <200 Mo (resp. <260 Mo) are included. The observations are from various sources which are discussed in De Donder and Vanbeveren (2004).

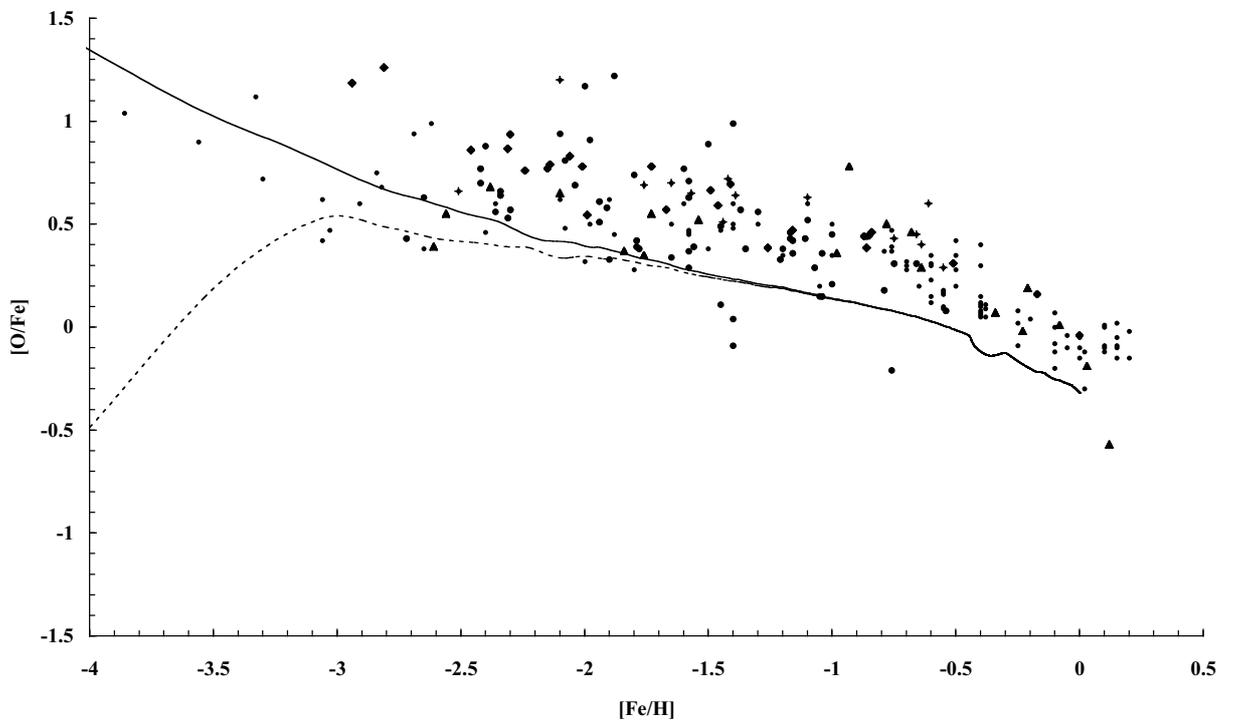

Figure 2: Similar as Fig1 but for $^{16}$O.

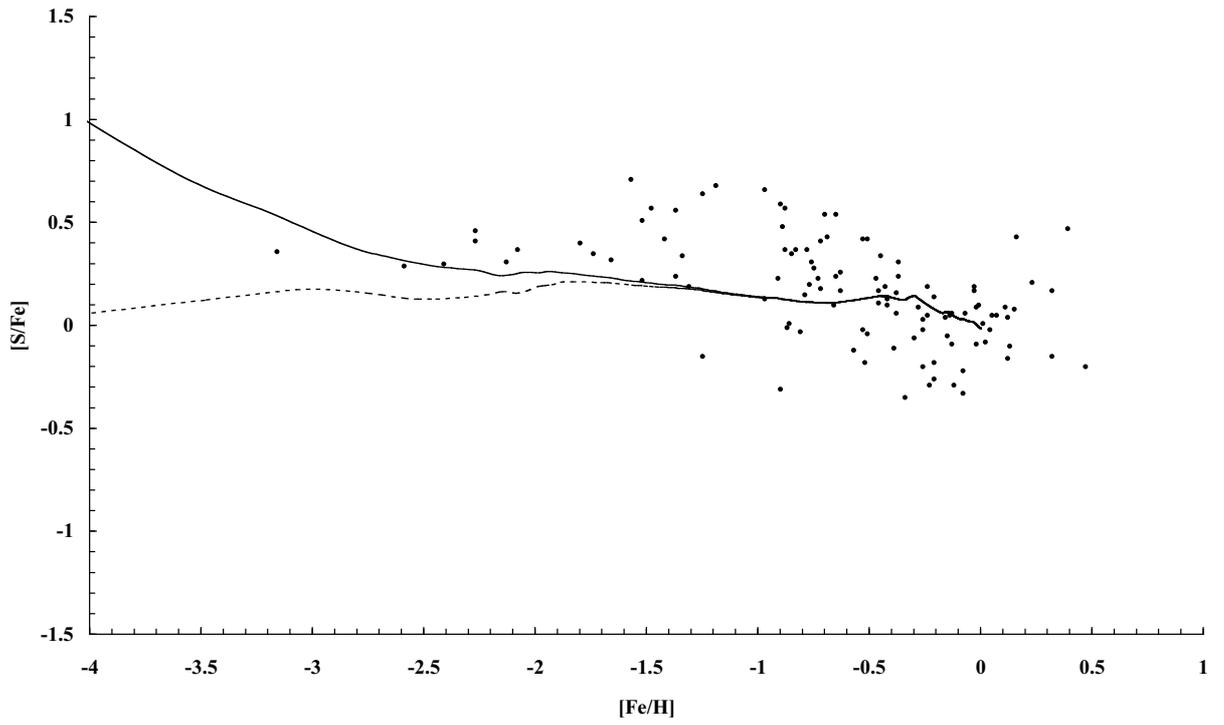

Figure 3: Similar as Fig1 but for $^{32}$S.

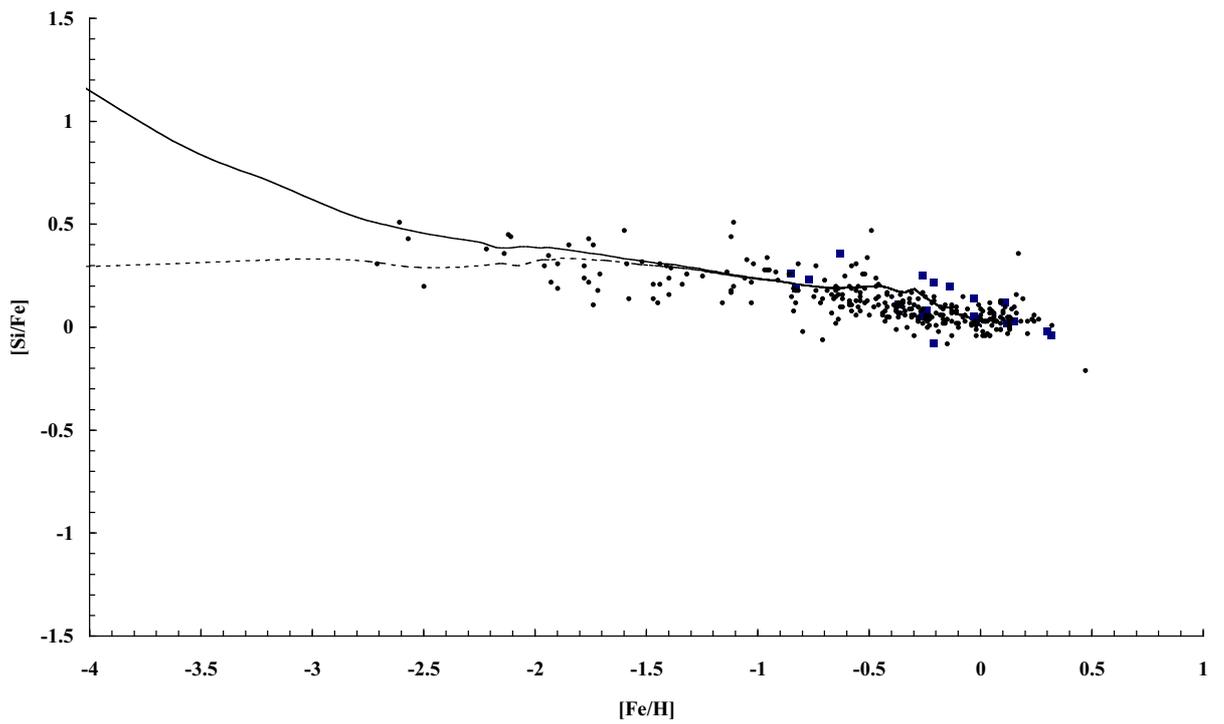

Figure 4: Similar as Fig 1 but for $^{28}$Si.

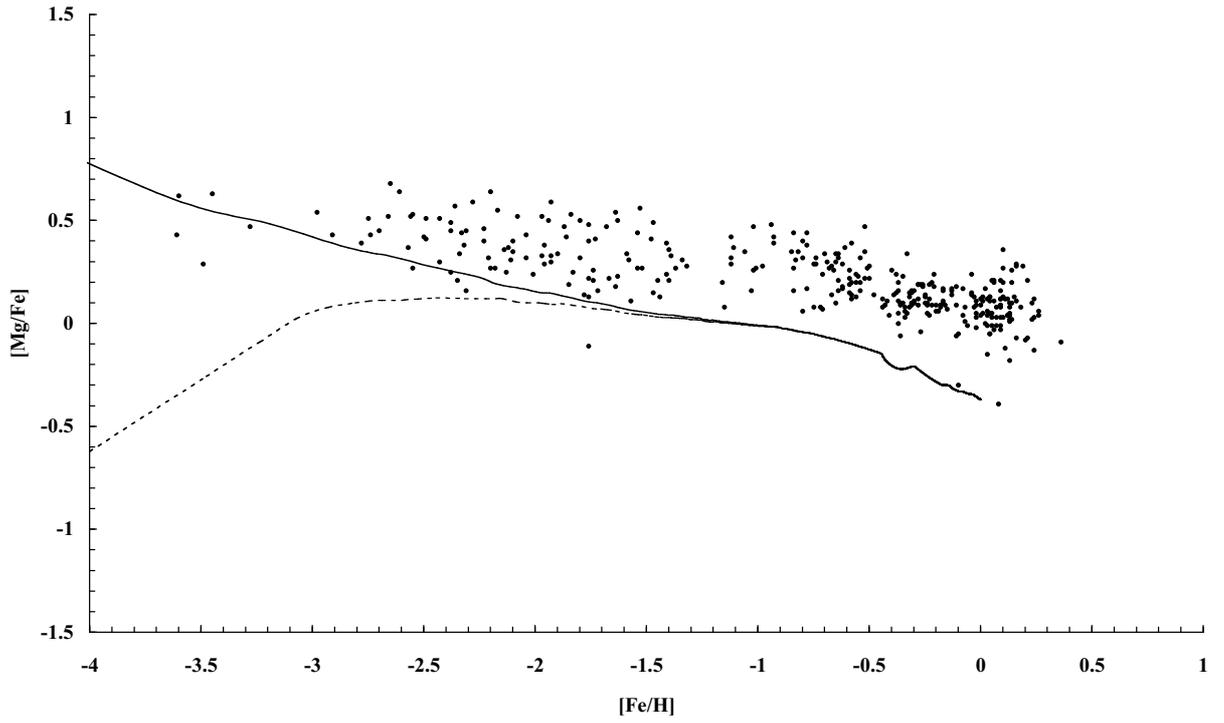

Figure 5: Similar as Fig. 1 but for $^{24}$Mg.

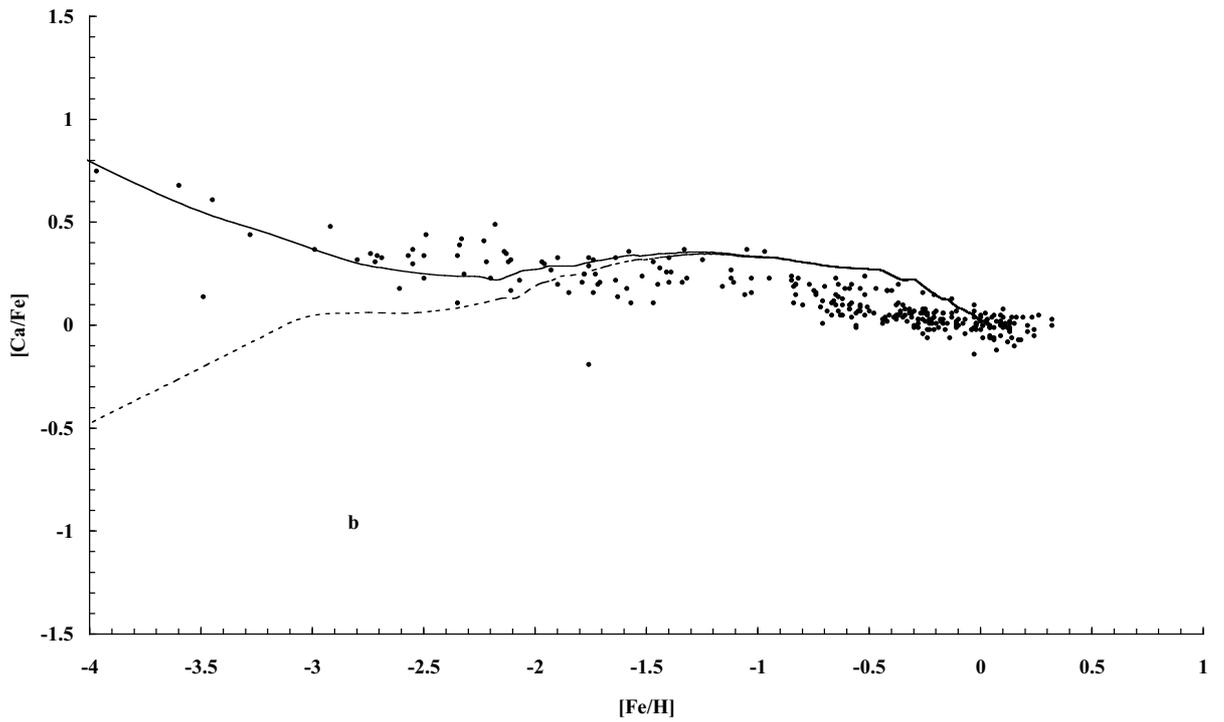

Figure 6: Similar as Fig. 1 but for Ca.

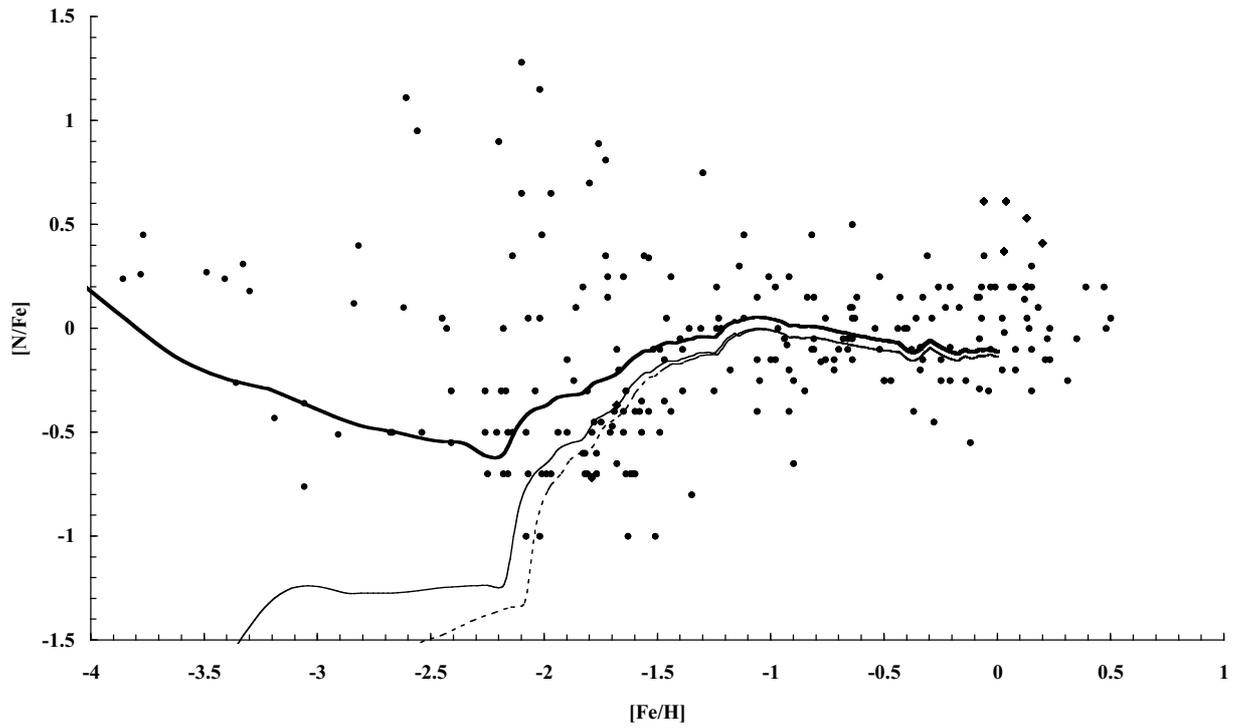

Figure 7: The observed (dots) versus the theoretically predicted temporal evolution of $^{14}$N. The dashed line is a CEM simulation without very massive stars. The thin line corresponds to the case where very massive single stars with a mass < 200 Mo are included but no very massive binaries; the thick line corresponds to the case where we account for a 20% very massive binary frequency. The observations are from various sources which are listed and discussed in De Donder and Vanbeveren (2004).

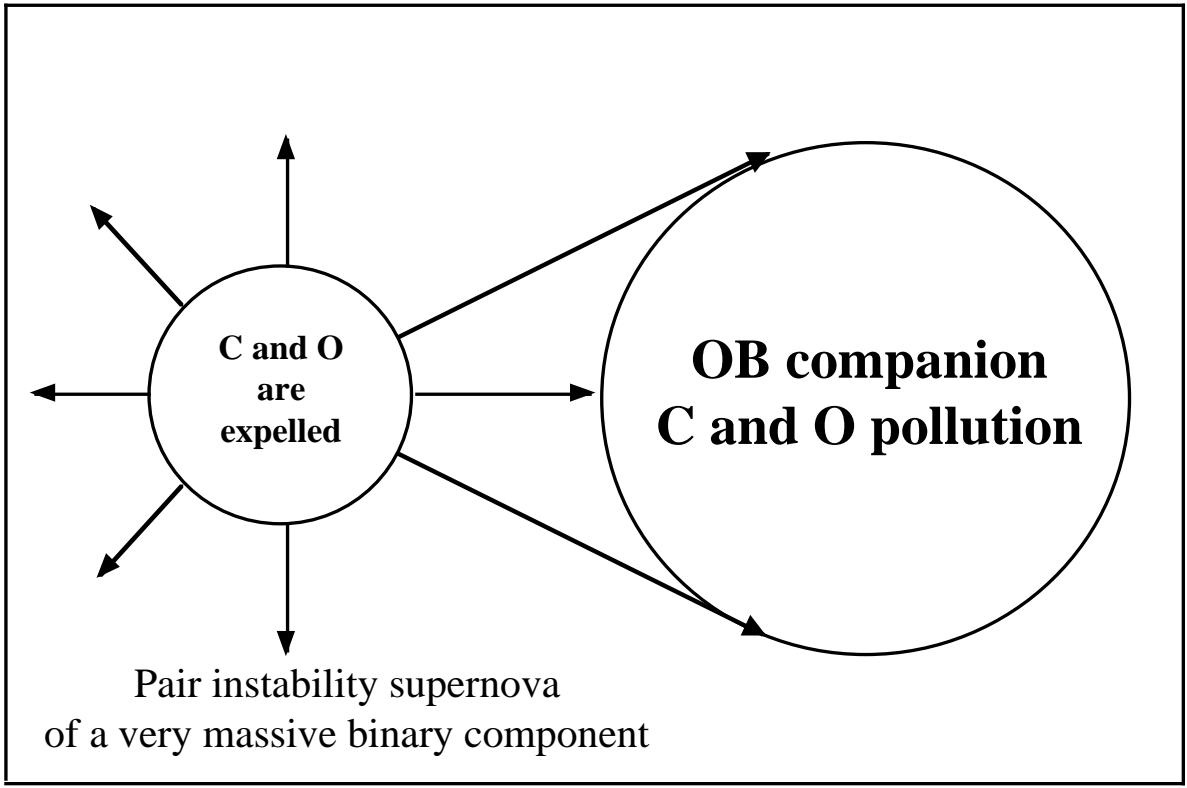

Figure 8: The very massive primary explodes as a pair instability SN and pollutes its companion.

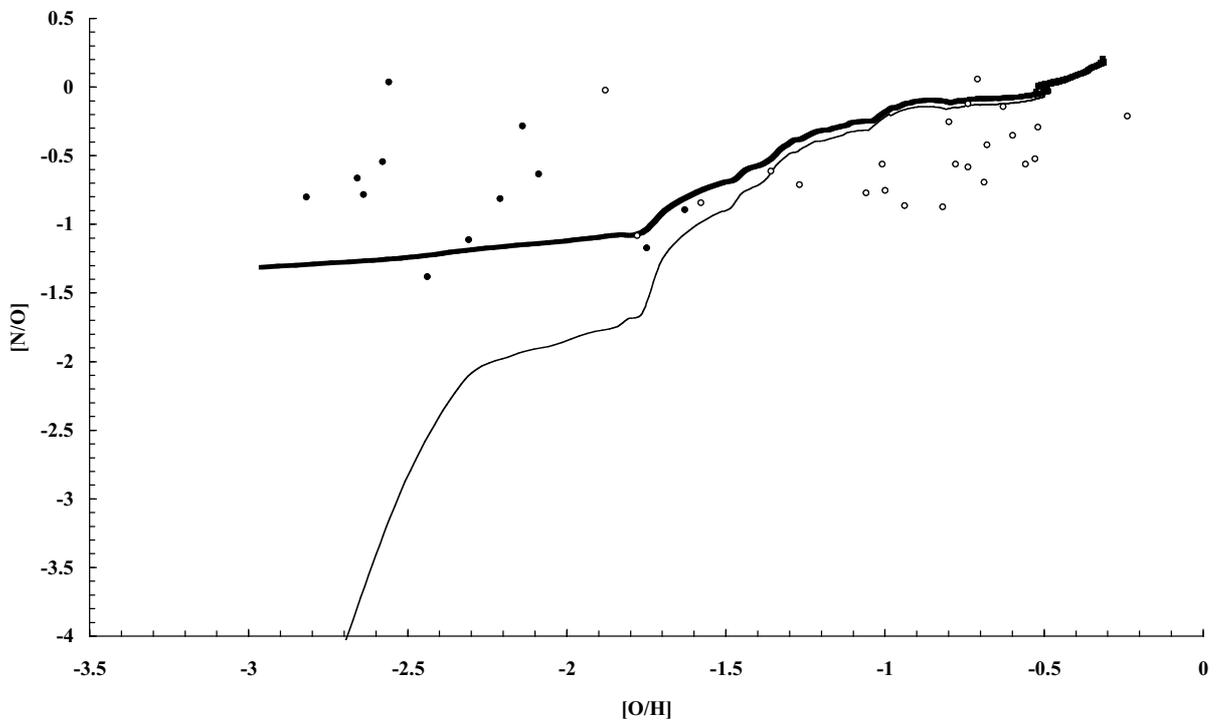

Figure 9: The theoretically predicted evolution during the early phases of the Galaxy of [N/O] as a function of [O/H]. The observations are from Israelian et al. (2004) and Spite et al. (2005).

The result shown in Figure 7 illustrates the conclusion that very massive single stars truncated at 200 Mo are unable to solve the $^{14}$N discrepancy. Could very massive interacting binaries provide a way out? (remind that stellar dynamics in N-body environments predicts that most of the massive and very massive stars are binary members, section 2.) Let us consider a typical very massive binary consisting of a 90 Mo He star (the post Roche lobe overflow remnant of a 180 Mo very massive star) with a 140 Mo companion. Low metallicity stars with a mass larger than 140 Mo have nearly equal core hydrogen burning timescales. This means that when the He star explodes, our 140 Mo is at the beginning of its core helium burning phase. Defining R as the radius of the companion, A as the binary separation, it is easy to understand that roughly $\frac{1}{4} 90 \frac{R^2}{A^2}$ Mo of the supernova matter (with a chemical composition which can be deduced directly from the yields published by Heger and Woosley and discussed above) will be accreted by the companion (Figure 8 illustrates what happens). The evolutionary consequences are very similar to those during the Roche lobe overflow of a massive binary where nuclearly processed matter is transferred from the loser towards the gainer. The accreted matter has a molecular weight µ larger than the µ of the underlying layers which is an unstable configuration that initiates thermohaline mixing (Kippenhahn et al., 1980). Thermohaline mixing is a very rapid process and is capable to mix all layers outside the He-core (Braun, 1997). For the purpose of the present paper we treat the process as an instantaneous one and we mix the accreted mass in a homogeneous way with all the layers outside the He core. The foregoing instantaneous mixing process is strengthened by the fact that very massive core helium burning stars are supra-Eddington and most of the layers outside the He-core are convectively unstable. The core helium burning evolution of the companion after this mixing process is quite interesting. Depending on the period of the binary (thus on the value of A) the layers outside the He-core typically contain 0.3-0.5 % (mass fraction) of $^{12}$C (and about a factor of 10 more of $^{16}$O), thus also the layers where hydrogen burning takes place. The CN cycle in these layers rapidly transforms the available $^{12}$C into $^{14}$N which is ejected as primary nitrogen during the SN explosion. Our stellar evolutionary calculations for the 140 Mo companion reveal that about 0.02-0.05 Mo (depending on the binary period, thus on A) of $^{14}$N are ejected, large enough to hope that it could solve the discrepancy between the observed and the theoretically predicted galactic temporal evolution of this element. To calculate whether or not very massive close binaries could explain the $^{14}$N discrepancy, we adapted our standard CEM and we assumed that 20% of all very massive stars are very massive close binaries which typically eject 0.05 Mo of $^{14}$N, as discussed in the previous section and we assumed that the percentage of very massive binaries remains constant for Z < 10$^{-5}$. It can readily be understood that the $^{14}$N yield predicted by the binary model discussed above scales lineary with the adopted binary frequency. A CEM where a 10\% (resp. a 50\%) very massive binary frequency is adopted would predict a $^{14}$N yield that is a factor 2 lower (resp. a factor 2.5 larger) that the simulation with a 20% very massive binary frequency. Figure 7 also compares the observations with our prediction. We conclude:

> *within the uncertainties of the content of a population of very massive stars during the early evolution of our Galaxy, very massive close binaries can produce the observed early temporal evolution of $^{14}$N, if the binary component is polluted by the pair instability supernova ejecta of its companion and explodes.*

Since the amount of accreted matter is small, the very massive binary process discussed above hardly affects the chemical evolution of the elements different from $^{14}$N.

To further illustrate the effect of very massive binaries, Figure 9 shows the [N/O]-[O/H] predictions and we compare them with observations of Israelian et al. (2004) and Spite et al.

(2005). As can be noticed the significant improvement when very massive binaries are included is also visible here. Of course the uncertainties mentioned in the conclusion above are very large, but the fact that a CEM, which includes the possible consequences of the process discussed above, produces a $^{14}$N slope which corresponds to the observed one, makes it worthwhile to explore the process in the futurein more detail.

**Acknowledgement**

The authors are very grateful to an unknown referee for very valuable remarks and comments that significantly improved the content of the paper.

**References**

Abel, T., Bryan, G.L., Norman, M.L.: 2002, Science 295, 93.

Akerman, C.J., Carigi, L., Nissen, P.E., Pettini, M., Asplund, M.: 2004, A&A 414, 931.

Argast, D., Samland, D., Thieleman, F.-K., Qian, Y.-Z.: 2004, A&A 416, 997.

Ballero, S., Matteucci, F., Chiappini, C.: 2006, NewA 11, 306.

Bally, J.: 2002, ASPC 267, 219.

Bally, J. and Zinnecker, H.: 2005, ApJ. 129, 2281.

Bate, M.R., Bonnell I.A., Bromm, V.: 2003, MNRAS 339, 577.

Beech, M., Mitalas, R.: 1994, ApJS 95, 517.

Bonnell, I.A., Bate, M. R., Zinnecker, H.: 1998, MNRAS 298, 93.

Bonnell, I.A., Bate, M.R., Vine, S.G.: 2003, MNRAS 343, 413.

Bonnell, I.A., Bate, M.R.: 2002, MNRAS 336, 659.

Bonnell, I.A.: 2002, ASPC 267, 193.

Bonnell, I.A.: 2005, in Massive Stars in Interacting Binaries, eds. A. Moffat and N. St.-Louis (in press).

Boss, A.P.: 1996, ApJ 468.231.

Braun, H.: 1997, Ph.D. thesis, Ludwig-Maximilians-Univ. München.

Bromm, V., Larson, R.B.: 2004, ARA&A 42, 79.

Chiappini, C., Matteucci, F., Gratton, R.: 1997, ApJ 477, 765.

Chiappini, C., Matteucci, F., Ballero, S.K.: 2005, A&A 437, 429.

Chiappini, C., Hirschi, R., Meynet, G., Ekström S., Maeder, A., Matteucci, F.: 2006 A&A (in press)(Astro-ph/0602459).

Clarke, C.J., Bonnell, I.A., Hillenbrand, L.A.: 2000, in Protostars and Planets IV (Book - Tucson: University of Arizona Press; eds Mannings, V., Boss, A.P., Russell, S. S.), p. 151.

De Donder, E., Vanbeveren, D.: 2003, New Astron. 8, 415.

De Donder, E., Vanbeveren, D.: 2004, New Astron. Rev. 48, 861.

de Wit, W. J., Testi, L., Palla, F., Vanzi, L., Zinnecker, H.: 2004, A&A 425, 937.

Fryer, C. L.: 1999, ApJ 522, 413.

Fryer, C.I., Woosley, S.E., Heger, A.: 2001, ApJ 550, 372.

Hachisu, I., Kato, M., Nomoto, K.: 1996, ApJ. 470, L97.

Hachisu, I., Kato, M., Nomoto, K.: 1999, ApJ. 522, 487.

Heger, A., Woosley, S.E.: 2002, ApJ 567, 532.

Hogeveen, S. J.: Pp\&SS 196, 299.

Iben, Jr. I., Tutukov, A. V.: 1984, ApJ. SS. 54, 335.

Israelian, G., Ecuvillon, A., Rebolo, R. et al.: 2004, A&A 421, 649.

Kippenhahn, R., Ruschenplatt, G., Thomas, H.C.: 1980, A&A 91, 175.

Klessen, R.S., Burkert, A.: 2000, ApJS 128, 287.

Kogut, A., Spergel, D. N., Barnes, C.: 2003, ApJS 148, 161.

Kudritzki, R.P., Pauldrach, A., Puls, J., Abbott, D.C.: 1989, A&A 219, 205.

Lada, C.J., Lada, E.: 2003, ARA&A 41, 57.

Larson, R.B.: 2000, in Star Formation from the Small to Large Scale, ed. F. Favata, A.A. Kaas. A. Wilson (ESP SP-445; Noordwijk: ESA), 13.

Larson, R.B.: 2003, Rep. Prog. Phys. 66, 1651.

Marigo, P., Chiosi, C., Kudritzki, R.-P.: 2003, A&A 399, 617.

Marigo,P., Girardi, L., Chiosi, C., Wood, P.R.: 2001, A&A 371, 152.

Maeder, A., Meynet, G.: 2004, A&A 422, 225.

Meynet, G., Maeder, A.: 2002, A&A 390, 561.

Meynet, G., Ekstrom, S., Maeder, A.: 2005, A&A (in press).

Penny, L.R.: 1996, PhD Thesis, Georgia State University.

Portegies Zwart, S. F., Makino, J., McMillan, S. L. W., Hut, P.: 1999, A&A 348, 117.

Portegies Zwart, S. F., McMillan, S. L.W.: 2002, ApJ. 576, 899.

Prantzos, N.: 2005, in Nuclei in the Cosmos VIII (Eds. L. Buchmann et al.) to appear in NuclPhys A (in press).

Sokasian, A., Abel, T., Hernquist, L., Springel, V.: 2003, MNRAS 344, 607.

Sommer-Larson, J., Götz, M., Portinari, L.: 2003, ApJ 596, 478.

Spite, M., Cayrel, R., Plez, B. et al.: 2005, A&A 430, 655.

Spruit, H.: 2002, A&A 381, 923.

Stahler, S.W., Palla, F., Ho, P.T.P.: 2000, in Protostars and Planets IV, ed. V. Mannings, A, Boss and A. Russell (Tucson: Univ. Arizona Press), 327.

Tinsley, B.M.:1980, Fundam. Cosmic Phys. 5, 287.

Vanbeveren, D.: 1991, A&A 252, 159.

Vanbeveren, D., De Donder, E., Van Bever, J., Van Rensbergen, W., De Loore, C.: 1998, NewA 3, 443.

Webbink, R, F.: 1984, ApJ. 277, 355.

Wolfire, M.G., Cassinelli, J.P.: 1987, ApJ 319, 850.

Woosley, S.E., Weaver, T.A.: 1995, ApJSS 101, 181.

Yorke, H.W. and Sonnhalter, C.: 2002, ApJ 569, 846.

Zinnecker, H., Bate, M. R.: 2002, ASPC 267, 209.